# Tuning of LANSCE 805-MHz High-Energy Linear Accelerator with Reduced Beam Losses

Y.K. Batygin, F.E. Shelley, H.A. Watkins, LANL, Los Alamos, NM 87545, USA

*Abstract*

Suppression of beam losses is essential for successful operation of high-intensity linac. Historically, the values of the field amplitudes and phases of the side-coupled, 805-MHz LANSCE linac modules are maintained using a well-known delta-t tuning procedure. Transverse matching of the beam with accelerator is performed through adjustments of beam ellipses in 4D phase space with accelerator lattice using matching quadrupoles. Control of the beam-energy ramp along the length of a proton linear accelerator is required to improve tune of accelerator and decrease beam losses. Time-of-flight measurements of the H$^-$ beam energy are now being used to confirm and improve the overall control of the energy ramp along the linac. The time-of-flight method utilizes absolute measurements of beam energy using direct signals from beam at an oscilloscope, as well as the difference in RF phases measured as the beam passes installed delta-t pickup loops. A newly developed BPPM data acquisition system is used. Beam energy measurement along accelerator together with phase scans, klystrons output power control, and delta-t method, allow tuning of accelerator much more close to original design and reduce losses generated by linac. Details of the upgraded tuning procedure and results of tuning and operation are presented.

## INTRODUCTION

The LANSCE accelerator facility currently delivers beams to five experimental areas (see Fig. 1). Accelerator is equipped with two independent injectors for H$^+$ and H$^-$ beams, merging at the entrance of a 201.25 MHz Drift Tube Linac (DTL). The DTL performs acceleration up to the energy of 100 MeV. After the DTL, the Transition Region beamline directs a 100 MeV proton beam to the Isotope Production Facility, while the H$^-$ beam is accelerated up to the final energy of 800 MeV in an 805-MHz Coupled Cavity Linac (see Fig. 2). The H$^-$ beams, created by different time structure of a low-energy chopper, are distributed in the Switch Yard (SY) to four experimental areas: Lujan Neutron Scattering Center equipped with Proton Storage Ring (PSR), Weapon Neutron Research Facility, Proton Radiography Facility, and Ultra-Cold Neutron Facility. The Drift Tube Linac consists of 4 tanks, which amplitudes and phases are selected through absorber-collector phase scans. Coupled Cavity Linac includes 44 accelerating modules (modules 5-48). Modules 5-12 contain 4 accelerating tanks each, while modules 13-48 contain 2 tanks each. Amplitudes and phases of 805 MHz linac are selected using the "delta-t" method [1]. In order to independently control tuning of the machine, the time of flight method for beam energy measurement was added.

## BEAM LOSSES IN ACCELERATOR

Beam losses in the LANSCE accelerator are mostly determined by the most powerful 80-kW H$^-$ beam, injected into Proton Storage Ring. Beam losses are controlled by various types of loss monitors. The main control is provided by Activation Protection (AP) detectors, which are one-pint size cans with a photomultiplier tube immersed in scintillator fluid. AP detectors integrate the signals and shut off the beam if the beam losses around an AP device exceed 100 nA of average current. The same devices are used as beam loss monitors (LM), where the signal is not integrated and therefore one can see a real-time of beam loss across the beam pulse. Within operation period, sensitivity of certain AP devices might be decreased in order to avoid interruption of operation due to single AP device.

Another type of loss monitor are Ion Chamber (IR) detectors, which are used in the high energy transport lines (Line D, PSR, 1L, WNR). They are usually located in parallel with Gamma Detectors

(GDs) that feeds into the Radiation Safety System. An advantage of the IRs is that they do not saturate at high loss rates like the AP devices. The third type of beam loss monitor are Hardware Transmission Monitors (HWTM). The HWTM system measures the beam current losses between current monitors and can limit beam current to a value at one current monitor.

Distribution of beam losses along the accelerator facility are presented in Figs. 3, 4. Typical averaged beam losses along the linear accelerator are $2 \times 10^{-3}$ which corresponds to loss rate of $3 \times 10^{-6}$ m$^{-1}$, or 0.2 W/m. In the high-energy beamlines (HEBT), the total beam losses are around $4 \times 10^{-3}$ which corresponds to a loss rate of $2 \times 10^{-5}$ m$^{-1}$, or 1.6 W/m. Higher beam losses in the HEBT are explained by smaller transverse acceptance and the dispersive nature of the beamlines, which generates additional losses due to longitudinal (energy) tails in the beam.

Beam losses in high-energy part of accelerator facility are sensitive to selection of amplitudes and phases of 805-MHz linac. Random errors in RF amplitude and phase result in increase of amplitude of longitudinal oscillations. Increase of amplitude of relative momentum spread in a sequence of $N_a$ accelerating sections with relative error in RF amplitude $\delta E_o / E_o$, and error in RF phase $\delta \psi$ is estimated as [2]:

$$<\delta(\frac{\Delta p}{p})>=\sqrt{\frac{N_a}{2}[<\Delta g>^2 +\left(\frac{qE\lambda}{mc^2}\right)\frac{|\sin\varphi_s|}{2\pi\beta\gamma^3}<\delta\psi>^2]}, \quad (1)$$

where

$$<\Delta g>=\frac{eE\lambda\cos\varphi_s}{mc^2\beta_N}\sqrt{<\frac{\delta E_o}{E_o}>^2 + tg^2\varphi_s <\delta\psi>^2}, \quad (2)$$

$E = E_o T$ is the accelerating gradient, $E_o$ is the average field in the tank, $T$ is the transit time factor, $mc^2$ is the ion rest energy, $\lambda$ is the RF wavelength, $\beta_N$ is the average effective beam velocity in linac, and $\varphi_s$ is the synchronous phase. Substitution of 805-MHz linac parameters into Eqs. (1) - (2) gives an estimation of increase of momentum spread due to RF field errors:

$$<\delta(\frac{\Delta p}{p})>=\sqrt{\frac{N_a}{2}(1.5\cdot 10^{-7} <\frac{\delta E_o}{E_o}>^2 +4.6\cdot 10^{-6} <\delta\psi>^2)}. \quad (3)$$

Momentum spread of the 800-MeV beam is measured by wire scanner LDWS03 located in a high-dispersive point of Line D (see Fig. 5). Momentum spread, $\Delta p / p$, is determined through measured beam size $R_x$ as

$$\frac{\Delta p}{p} = \frac{\sqrt{R_x^2 - \beta_x (4 \ni_{rms})}}{\eta}, \quad (4)$$

where dispersion of the lattice at that point is $\eta$ = 4.8798 m and horizontal beta-function $\beta_x$ =1.11236 cm /mrad. For typical unnormalized 800 - MeV beam emittance $\ni_{rms}$ = 0.04 π cm mrad, the usual beam size of $R_x$ = 5 mm is mostly determined by beam momentum spread of $\Delta p / p$ = $8.13\cdot 10^{-4}$. The estimation of increase of momentum spread, Eq. (3), indicates that for instability of the RF field amplitude $<\delta E_o / E_o> \approx 1\%$ and that of phase $<\delta\psi> \approx 1^o$, increase of momentum spread of the beam is around $<\delta(\Delta p / p)> \approx 1.7\cdot 10^{-4}$, which is a significant addition to regular momentum spread of the beam. Improper tune of accelerator results in appearance of momentum tails in beam distribution (see Fig. 6), which generates significant beam spill. Precise selection of amplitudes and phases of high-energy linac are important to minimize beam losses in accelerator facility.

## DELTA-T TUNING PROCEDURE

Historically, selection of amplitudes and phases of accelerating modules of 805-MHz LANSCE linac were performed using classical "delta-t" procedure [1]. This turn-on method is based on measurement of difference in time of flights between 2 pairs of pickup loops when accelerating module is off and on, and compare this differences with design differences. Let $t_{AB}$ and $t_{AC}$ be the time-of-flight of the beam centroid from location A to B, and A to C, respectively (see Fig. 7). Then, the change in these values when accelerating module is switched from off to on are

$$t_B = t_{AB,off} - t_{AB,on}, \qquad t_C = t_{AC,off} - t_{AC,on}. \tag{5}$$

Deviation of values $t_B$, $t_C$ from design values are expressed as [1]

$$\Delta t_B = -\frac{D_{AB}}{mc^3(\beta\gamma)_A^3}\Delta W_A - \frac{\Delta\varphi_B - \Delta\varphi_A}{\omega} - \frac{D_1}{mc^3}[\frac{\Delta W_A}{(\beta\gamma)_A^3} - \frac{\Delta W_B}{(\beta\gamma)_B^3}], \tag{6}$$

$$\Delta t_C = \Delta t_B - \frac{D_2 - D_1}{mc^3}[\frac{\Delta W_A}{(\beta\gamma)_A^3} - \frac{\Delta W_B}{(\beta\gamma)_B^3}], \tag{7}$$

where $\Delta W_A$, $\Delta W_B$, $\Delta\varphi_A$, $\Delta\varphi_B$ are deviation from design energy and RF phase at the entrance and at the exit of accelerating module. When accelerator is tuned according to design, $\Delta t_B = 0$, $\Delta t_C = 0$. Delta-t is an automatic procedure, which performs search for amplitude and phase of each module to minimize values of $\Delta t_B$, $\Delta t_C$. The search is usually terminates when values of $\Delta t_B$, $\Delta t_C$ becomes less than 0.01 nsec. If initial values of beam energy, RF amplitude and phase are selected to be close to design values, the method provides quick convergence (see Fig. 8).

Design values of $t_B$, $t_C$ are obtained from idealized computer model of the linac, assuming there are no RF phase shifts between tanks in a module, and that the RF amplitude is the same in each tank in module. Tank-to-tank amplitude or phase errors larger than 1% or 1°, respectively, can significantly affect results of the delta-t tuning [1]. The accelerating field is also known to have some variation within each tank, which is not taken into account in computer model. Such effects result in deviation of accelerator tune from deign. Delta-t method uses only a small phase range (~ 10°) for tuning, because the procedure is based on a linear model. Delta-t procedure does not provide information about actual energy before and after accelerating module. Deviation of beam energy from design at the entrance and at the exit of accelerating module, $\Delta W_A$, $\Delta W_B$, are calculated from simplified linear model, Eqs. (6) - (7), assuming that values are small ($\Delta W / W \sim 10^{-3}$). If deviation of input energy from design value is not small, and / or initial values of amplitudes and phases are not close to design values, the delta-t search process might be unstable, and does not converge (see Fig. 9). Therefore, values of amplitude and phase as well as input energy before delta-t search have to be selected as close as possible to design values.

Delta-t tuning procedure works well when particles perform significant longitudinal oscillations within RF tanks [3]. If longitudinal oscillations are "frozen", then values $\Delta t_B$, $\Delta t_C$ can be obtained with infinitely large number of combinations of amplitudes $E$ and synchronous phases $\varphi_s$. Suppose, combination ($E_1$, $\varphi_{s1}$) generates certain values $\Delta t_B$, $\Delta t_C$. If beam phase is changed, $\varphi_{s2} = \varphi_{s1} + \Delta\varphi$, then $\Delta t_B$, $\Delta t_C$ are also changed. But for cavity with negligibly small value of phase advance of longitudinal oscillations, this change can be compensated by another field $E_2$, which provides the same values of $\Delta t_B$, $\Delta t_C$, as for combination ($E_1$, $\varphi_{s1}$). Therefore, for "short" cavities, selection of amplitudes and phases is not a well determined process. Phase advance of longitudinal oscillation per module is [2]

$$\mu_{ol} = \sqrt{2\pi(\frac{qE\lambda}{mc^2})\frac{|\sin\varphi_s|}{(\beta\gamma)^3}}(\frac{L_t}{\lambda}), \tag{8}$$

where $L_t$ is the total module length. Accelerating gradient in linac is $E \approx 1$ MV/m, and wavelength $\lambda = 0.3726$ m. In the beginning of 805 MHz linac, at the energy of 100 MeV, $\varphi_s = -36°$, $\beta\gamma = 0.4767$, $L_t = 14.67$ m, $\mu_{ol} = 264°$. At the end of linac, where beam energy is 800 MeV, $\varphi_s = -30°$, $\beta\gamma = 1.558$, $L_t = 17.2$ m, $\mu_{ol} = 48°$. The value of longitudinal phase advance drops as or factor of 5.5 from 100 MeV to 800 MeV. It results in less accurate tune of the machine with energy. In order to control independently tuning process, the time-of-flight beam energy measurements and phase scans were added to tuning process. These methods are successfully used in other accelerator facilities for tuning purposes [4] - [7].

## BEAM ENERGY MEASUREMENT BY TIME OF FLIGHT

Energy measurement of the beam is performed through measurement of time of flight (TOF) of beam $t = L/\beta c$ between 2 pickup loops (see Figs. 7, 10), separated by distance $L$, where $\beta c$ is the beam velocity [8]. Time of flight can be represented as

$$t = N\frac{\lambda}{c} + \Delta t, \qquad (9)$$

where $N$ is the integer number of RF periods during time of flight, and $\Delta t$ is the fractional part of time of flight. The change of RF phase during time of flight is

$$\omega t = 2\pi N + \Delta\varphi, \qquad (10)$$

where $\Delta\varphi = 2\pi c \Delta t / \lambda$ is an actual measured fractional part of phase change. Expressing beam velocity from Eqs. (9), (10):

$$\beta = \frac{L}{\lambda(N + \frac{\Delta\varphi}{2\pi})}, \qquad (11)$$

the beam energy is determined as

$$W = mc^2 \left(\frac{1}{\sqrt{(1-\beta^2)}} - 1\right). \qquad (12)$$

The error in determination of beam velocity is

$$\frac{d\beta}{\beta} = \sqrt{\left(\frac{dL}{L}\right)^2 + \left(\frac{\delta(\Delta\varphi/2\pi)}{N + \frac{\Delta\varphi}{2\pi}}\right)^2}, \qquad (13)$$

where $dL/L$ is the error in measured distance between pickup loops, and $\delta(\Delta\varphi)$ is the error in phase difference measurement, while error in determination of energy is

$$\frac{dW}{W} = \gamma(\gamma+1)\frac{d\beta}{\beta}. \qquad (14)$$

Separation in possible values of energy due to ambiguity in velocity determination is ($N \gg 1$)

$$\frac{\Delta W}{W} \approx \frac{\gamma(\gamma+1)}{N}. \qquad (15)$$

Determination of beam energy requires knowledge of integer number of RF periods during time of flight. Direct observation of beam pulse trains (see Figs. 11, 12) allows measurement of absolute value of time of flight between two loops [4]. Beam velocity is determined as

$$\beta = \frac{L}{c[t - (\tau_{cable2} - \tau_{cable1})]}, \quad (16)$$

where $t$ is the observable time of flight at the scope, and $\tau_{cable1}$, $\tau_{cable2}$ are cable lengths measured in terms of time required for signal propagation from pickup loop to the scope. Error in determination of velocity is

$$\frac{d\beta}{\beta} = \sqrt{(\frac{dL}{L})^2 + (\frac{dt}{t - \Delta\tau_{cable}})^2 + [\frac{d(\Delta\tau_{cable})}{t - \Delta\tau_{cable}}]^2}, \quad (17)$$

where $\Delta\tau_{cable} = \tau_{cable2} - \tau_{cable1}$ is the cable length difference.

## BEAM ENERGY MEASUREMENT ALONG 805-MHZ LINAC

For beam energy measurement, we used existing delta-t loops distributed along the linac. Both methods, presented above, included preliminary calibration of cable lengths, which was done using beam with known energy. Accelerated bunches are separated by the distance of $\beta\lambda$, where $\beta$ is the particle velocity, and $\lambda = 1.48965$ m is the wavelength of 201.25 MHz DTL linac. Final beam energy of the linac 795.46 MeV is determined through operation of LANSCE Proton Storage Ring, which has circumference of 90.26 m and operational frequency of 2.792424 MHz. Final pair of delta-t loops after linac (loops 48 – SY) is separated by the distance of $L_{48-SY} = 20.029$ m. Measured time of flight $t_{48} = 146.6$ nsec between those delta-t loops with known beam velocity $\beta_{48} = 0.84073$ provides beam-based measured difference in cable length

$$\tau_{SY} - \tau_{48} = t_{48} - \frac{L_{48-SY}}{\beta_{48}c} = 67.13 \text{ ns}. \quad (18)$$

For measurement of beam energy after every accelerating module, each module was subsequently delayed and TOF between loops 48-SY was determined. Measurements started from the last module 48 and performed until module 13, unless the signal from drifting beam was observed. Using Eqs. (12), (16), the particle velocity and beam energy were determined.

The similar technique was used to measure beam energy in the beginning of 805-MHz linac. Beam with known energy of 100 MeV after DTL was used to calibrate cable length difference of delta-t loops after modules 11-12. Using distance between loops $L_{11-12} = 16.976$ m, velocity of 100 MeV particles $\beta_{100\,MeV} = 0.42799$, and measured TOF $t_{100\,MeV} = 196.15$ ns, the beam-based measured difference in cable length of loops 11-12 is:

$$\tau_{12} - \tau_{11} = t_{100\,MeV} - \frac{L_{11-12}}{\beta_{100\,MeV}c} = 63.84 \text{ ns}. \quad (19)$$

Turning on subsequently modules 5-11, we measured TOF and beam energy after modules 5-11. The final determination of energy after module 12 was performed using measured beam energy after module 11, $E_{11} = 196.318$ MeV, and delta-t loops after modules 12-13, which gave the value of cable difference between loops 12-13:

$$\tau_{13} - \tau_{12} = t_{196.3 MeV} - \frac{L_{12-13}}{\beta_{196.3 MeV} c} = 58.18 \text{ ns} \quad . \tag{20}$$

Typical error in determination of time of flight in absolute measurements is $dt = \Delta\tau_{cable} = 0.1$ ns. Distances between delta-t loops are known with relative error of $dL/L = 5 \cdot 10^{-5}$. The error in absolute determination of energy is

$$\frac{dW}{W} \simeq (1.1 - 1.8) \cdot 10^{-3} \gamma(\gamma + 1). \tag{21}$$

Absolute energy measurements were used as a reference for more precise energy determination using measurement of RF phase difference in delta-t loops.

A newly developed BPPM data acquisition system to control the 3D position of the beam centroid (x, y, phase) was used [9]. As in previous method, we started measurements from the end of the linac using 48-SY delta-t loops. Measured beam phase difference between loops $(\varphi_{SY} - \varphi_{48})_{795 MeV} = 192°$ together with expected difference in beam phase

$$\Delta\varphi_{expected\_48\_SY} = 2\pi[\frac{L_{48-SY}}{\beta_{48}\lambda} - INT(\frac{L_{48-SY}}{\beta_{48}\lambda})] = 357.3°, \tag{22}$$

provides phase correction due to difference in cable lengths

$$\Delta\varphi_{corr\_48\_SY} = \varphi_{expected\_48\_SY} - (\varphi_{SY} - \varphi_{48})_{795 MeV} = 165.3°. \tag{23}$$

Using the RF phase difference measurement in loops 48-SY after each module, $(\varphi_{SY} - \varphi_{48})$, and known value of integer RF periods for each value of energy from absolute measurements, $N$, the beam energy was determined after each module 13-48:

$$\beta = \frac{L_{48-SY}}{\lambda(N + \frac{\varphi_{SY} - \varphi_{48} + \Delta\varphi_{corr\_48\_SY}}{2\pi})}. \tag{24}$$

In order to perform beam energy measurement in modules 5-11, the calibration of 11-12 pair of delta-t loops was done using 100 MeV beam after DTL. Measured difference in phases $(\varphi_{12} - \varphi_{11})_{100 MeV} = 182.52°$ resulted in correction due to cable difference between modules 11-12:

$$\Delta\varphi_{corr\_11\_12} = 2\pi[\frac{L_{11\_12}}{\beta_{100 MeV}\lambda} - INT(\frac{L_{11\_12}}{\beta_{100 MeV}\lambda})] - (\varphi_{12} - \varphi_{11})_{100 MeV} = 42.95°. \tag{25}$$

The same method was used to calibrate delta-t loops 12-13 using 196.3 MeV beam after module 11:

$$\Delta\varphi_{corr\_12\_13} = 2\pi[\frac{L_{12\_13}}{\beta_{196.3 MeV}\lambda} - INT(\frac{L_{12\_13}}{\beta_{196.3 MeV}\lambda})] - (\varphi_{13} - \varphi_{12})_{196.3 MeV} = 94.23°. \tag{26}$$

Then, the beam velocity was determined though measured phase difference $\varphi_{loop2} - \varphi_{loop1}$ and correction values $\Delta\varphi_{corr}$:

$$\beta = \frac{L}{\lambda(N + \frac{\varphi_{loop2} - \varphi_{loop1} + \Delta\varphi_{corr}}{2\pi})}, \tag{27}$$

where value of $N$ was known from absolute TOF measurement. Error in determination of phase in considered method is $\delta(\Delta\varphi) \approx 1^o$. This method appears to be more accurate than the absolute energy measurement method. The estimated error is one order of magnitude smaller than that in absolute method:

$$\frac{dW}{W} \simeq (1.6 - 2.4) \cdot 10^{-4} \gamma(\gamma + 1). \tag{28}$$

Separation in energy due to ambiguity, Eq. (15), is two orders of magnitude larger than error, Eq. (28). Obtained results created basis for careful tuning of linear accelerator.

## TUNING OF 805-MHz LINAC

Beam adjustment to linac includes independent transverse and longitudinal matching of the beam to the structure. Focusing structure of 805-MHz linac is a FDO structure. Doublets of focusing-defocusing quadrupoles are located between accelerating tanks (see Fig. 2). Matching conditions of beam ellipses in front of linac are determined by the following combination of Twiss parameters: $\alpha_x = 0.337$, $\beta_x = 329$ cm, $\alpha_y = 1.24$, $\beta_y = 556$ cm [10]. Matching is performed using four quadruoles TRQM5-8 in Transition Region between DTL and CCL. Design of quadrupole structure assumes ramp of quadrupole gradient values between modules 5-13. Experience indicates that beam losses are sensitive to ramp of quadrupole values. Figures 13, 14 indicate significant reduction of beam losses in linac while empirical change of quadupole setups. After module 13 the values of quadrupole gradients are kept constant.

Tuning of 805-MHz accelerator sections is performed through selection of amplitudes and phases of each module. In addition to traditional delta-t method, we use phase scans, field control through klystron output power, and beam energy measurement. Preliminary selection of field amplitude in each module is determined from expected energy gain per module

$$\Delta W = qE_o T L_a \cos\varphi_s = qUT \cos\varphi_s, \tag{30}$$

where $U = E_o L_a$ is the voltage of accelerating module. Effective voltage

$$UT = \sqrt{P_{RF} R_{sh}} \tag{31}$$

is determined by RF power entering RF sections, $P_{RF}$, and shunt impedance of accelerating sections

$$R_{sh} = (ZT^2)L_a, \tag{32}$$

where $ZT^2$ are effective shunt impedance per unit length. Total power is a sum of RF power entering RF sections, $P_{RF}$, and power loss in waveguide connecting klystron and RF modules, and in bridges between accelerating tanks, $P_{loss}$:

$$P_{tot} = P_{RF} + P_{loss}. \tag{33}$$

The electrodynamics values of accelerating tanks of 805-MHz linac were determined in Ref. [11]. Table 1 contains design values of 805-MHz linac, which served as starting points for tuning process.

Phase scan is a first step in tuning procedure of each module. At that step, the beam phase at BPM after module is measured as a function of variable phase of accelerating module (see Fig. 15). The value of phase, where phase scan curve has a minimum, corresponds to maximum value of RF field (crest of RF field). The setup value of RF phase of module, $\varphi_s$, is selected as subtraction of the design value of synchronous phase from minimum curve phase value. Design values of $\varphi_s$ are changing linearly from -36° to -30° from module 5 to module 17, and then are kept constant as -30° until the end of linac. After each module, output energy was verified using technique written above, which served as an input energy for the next module. With carefully selected values of input energy, required klystron power, and RF phase, delta-t method provides quick convergence to tuned values (see Fig. 8).

Application of described method allows tune of machine much more close to design, than using delta-t method only. Figures 16, 17, and Table 2 contain results of beam energy measurement after application of described tuning procedure. Figure 18 illustrates energy gain per module after delta-t tuning and after improved combined tuning. As seen, in first case, deviation for design progressively increases with energy, while in second case energy gain per module is much more close to design. Figure 19 illustrates klystron power along accelerator after both tuning methods. Accelerator tune after delta-t tuning results in excessive power at certain klystrons. Particularly, klystron of Module 17 worked at the value of 1.1 MW, close to the peak limit of klystron. Application of combined tune results in equalization of klystron power along accelerator, which is close to design, avoiding peaks of required klystron power. Table 3 illustrates beam losses in high-energy part of accelerator facility after regular tune using only delta-t procedure, and with modified approach, using combination of phase scans, energy measurement, klystron power, and delta-t method. Application of improved tuning method resulted in significant (factor of 2) reduction of beam losses.

## SUMMARY

Time of flight beam energy measurements together with phase scans and klystron output power control were added to regular delta-t tuning of LANSCE 805-MHz high-energy linac. Developed approach allows effective control of beam energy ramp along linear accelerator. Method utilizes direct measurement of time-of-flight between pick-up loops, as well as measurement of difference in RF phases while beam crosses installed delta-t loops. Newly developed chassis to control 3D position of beam centroid is used. Application of combined method allows tune of accelerator much more close to design with significantly reduced beam losses generated by linac in high-energy part of accelerator facility, avoiding high-peaks in klystron power.

## ACKNOLEDGEMENTS

Authors are indebted to Lawrence Rybarcyk, Chandra Pillai, Prabir Roy, and Charles Taylor for help in performing of experiments and useful discussion of results.

## REFERENCES

[1] K.R.Crandall, "The Delta-T Tuneup Procedure for the LAMPF 805-MHz Linac", LANL Report LA-6374-MS, June 1976.
[2] I.M. Kapchinsky, "Theory of Resonance Linear Accelerators", Harwood, 1985.
[3] K.Crandall, D.Swenson, "Side Coupled Linac Turn-On Problem", Los Alamos Scientific Laborstory, Report MP-3-98, 1970.
[4] M.B. Popovic, T.L. Owens, T.K. Kroc, L.J. Allen and S.W. Schmidt, "Time-of-Flight Measurement of Absolute Beam Energy in the Fermilab Linac", Proceedings of the 1993 Particle Accelerator Conference (PAC 93), May 17-20, 1993 Washington D.C., p.1689.
[5] T.L. Owens, M.B. Popovoc, E.S. McCrory, C.W. Schmidt, and L.J .Allen, Particle Accelerators, 1994, Vol. 48, p.p.169-179.
[6] J. Galambos, A. Alexandrov, C. Deibele, S. Henderson, Proceedings of 2005 Particle Accelerator Conference, Knoxville, Tennessee, p.1491.
[7] A.Shishlo, A. Alexandrov, Proceedings of Hadron Beams 2008, Nashville, Tennessee, WGB06, p.203.
[8] P.Strehl, "Beam Instrumentation and Diagnostics", Springer, 2010.
[9] H.Watkins, J.D.Gilpatrik, R.McCrady, Proceedings of IBIC2014, Monterey, CA, USA, WEPD09, p. 655.
[10] G.Swain Proceedings of the Linear Accelerator Conference 1990, Albuquerque, New Mexico, USA, p. 367.
[11] G.Swain, "LAMPF 805-MHz Accelerator Structure Tuning and Its Relation to Fabrication and Installation", LA-7915-MS, 1979.

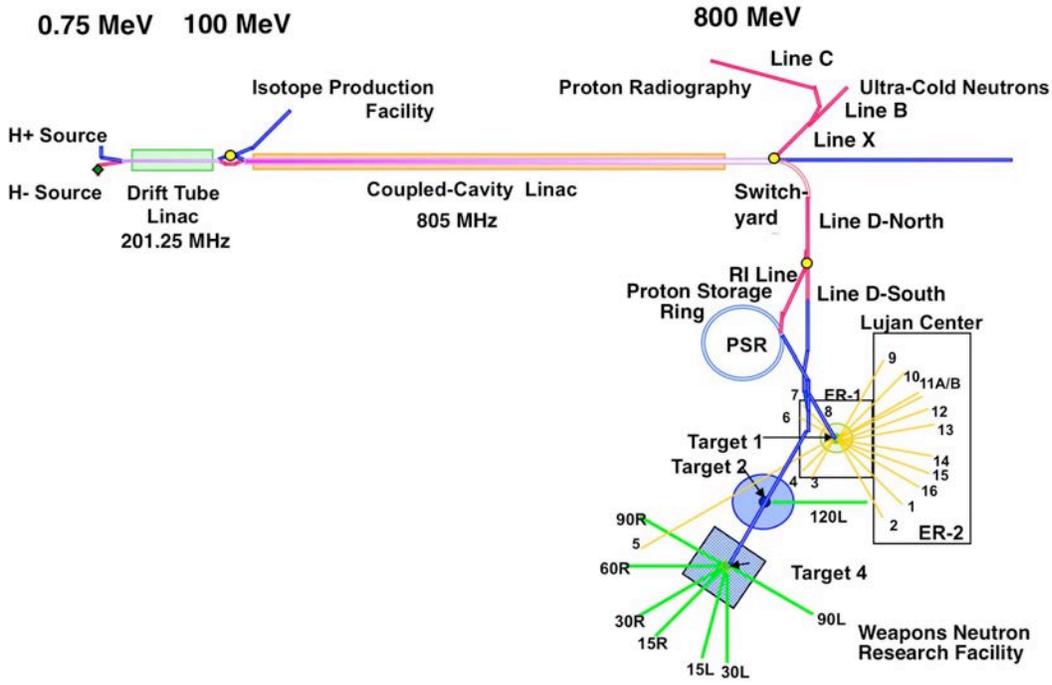

Figure 1: Layout of LANSCE Accelerator Facility.

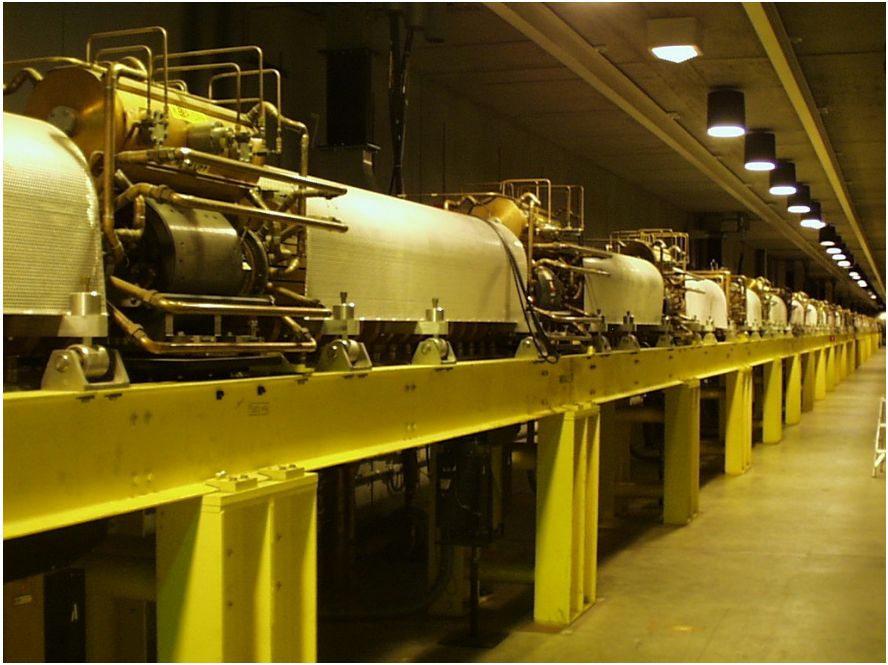

Figure 2: Accelerating tanks of 805 MHz Coupled-Cavity linac separated by quadrupole doublets.

Figure 3: Beam loss along linear accelerator.

Figure 4: Beam loss in high-energy beam transport.

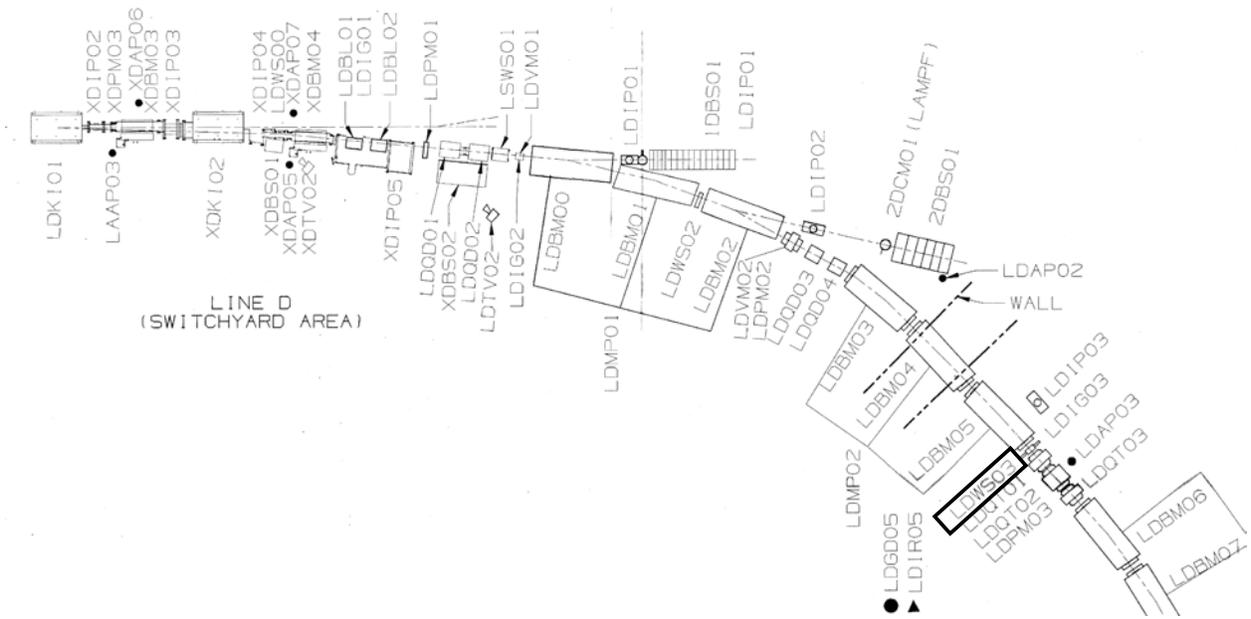

Figure 5: Location of beam spectrometer LDWS03 in Line D-North area.

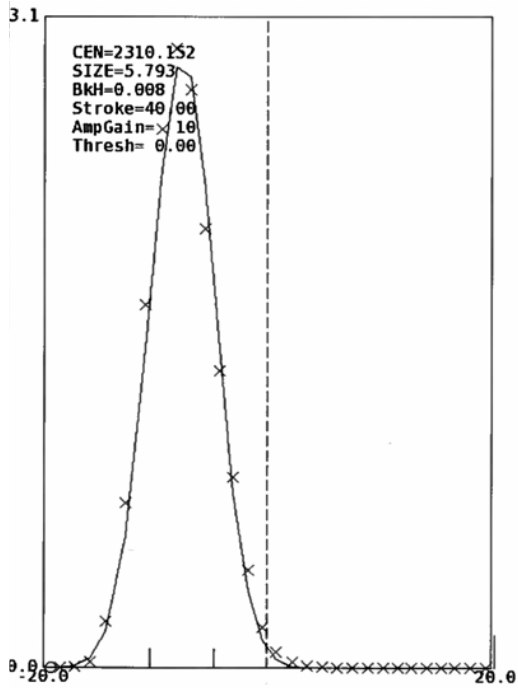
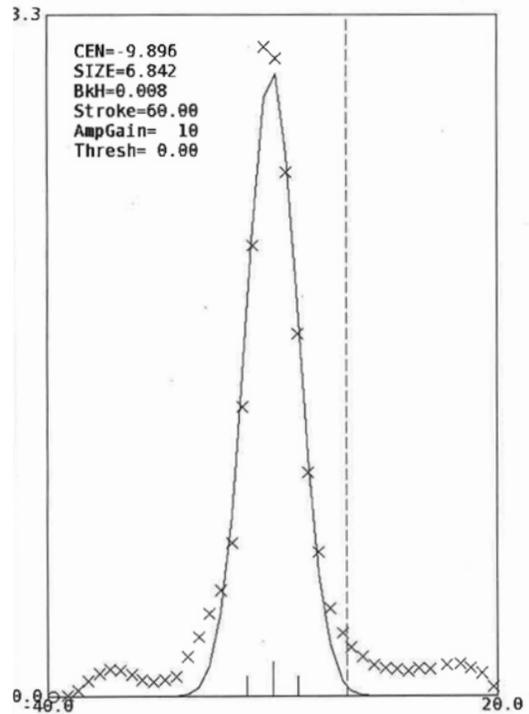

(a)                            (b)

Figure 6: Momentum spread of the beam measured by LDWS03 wire scanner: (a) properly tuned beam, (b) beam with momentum tails due to improper tune.

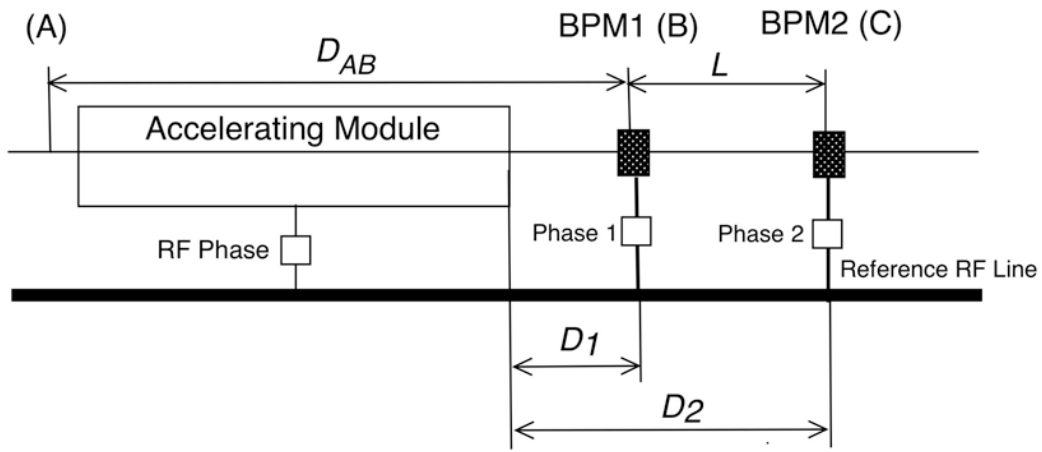

Figure 7: Setup for delta-t and beam energy measurements with two pickups loops BPM1, BPM2.

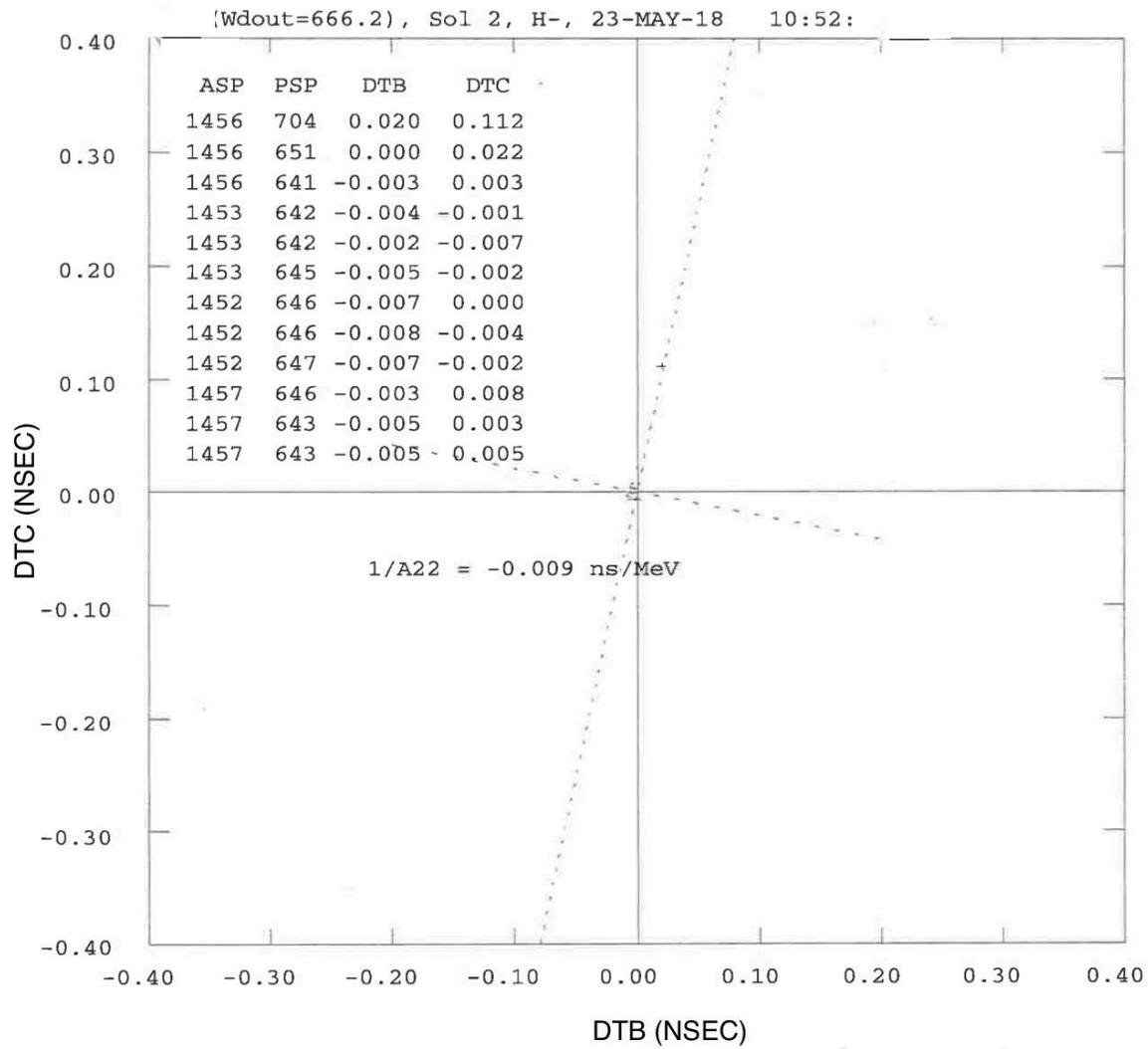

Figure 8: Output of delta-t program displaying search of amplitude (ASP) and phase (PSP) while minimizing values of $\Delta t_B$ (DTB) and $\Delta t_C$ (DTC).

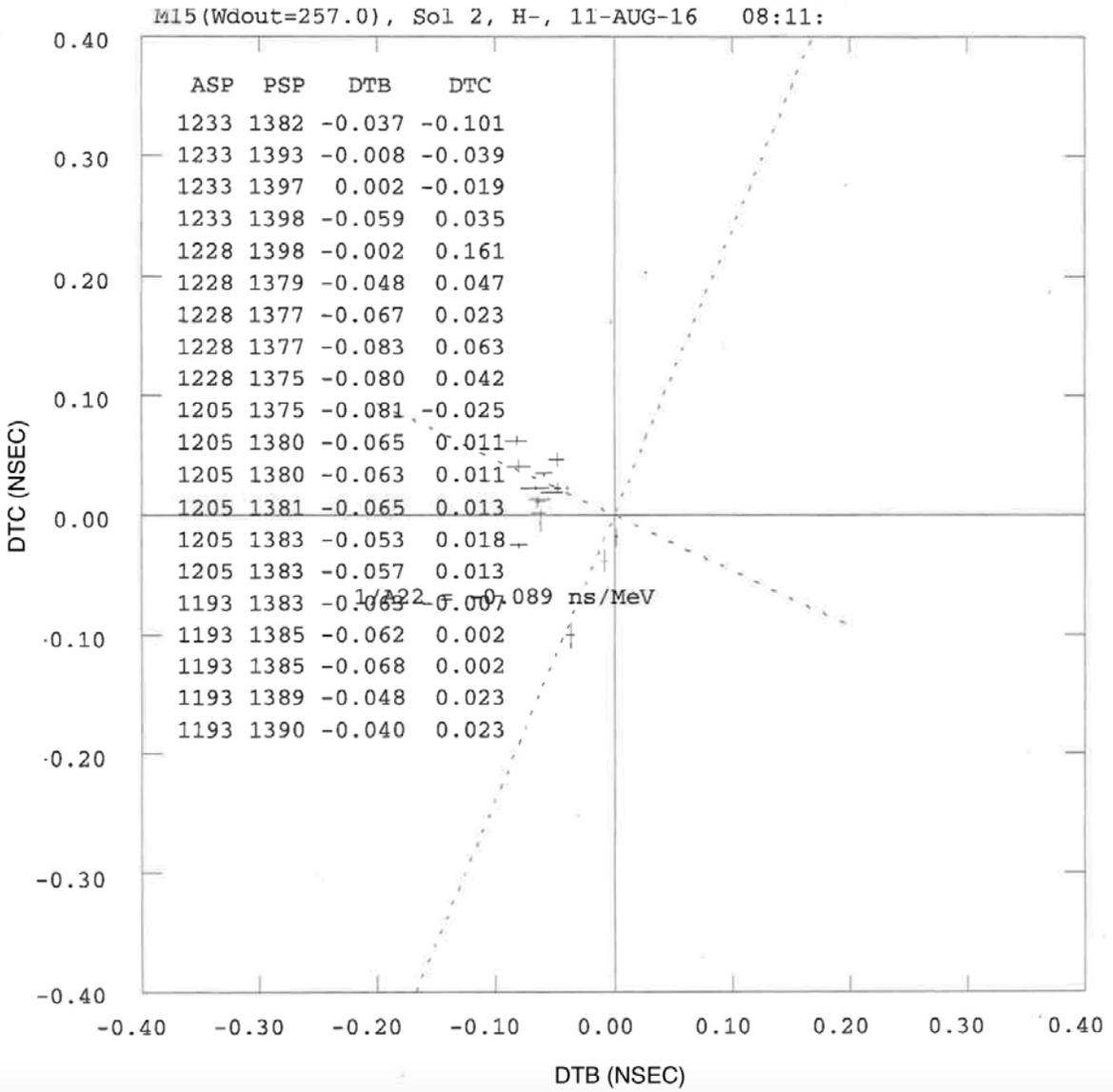

Figure 9: Example of unstable delta-t search of amplitude (ASP) and phase (PSP).

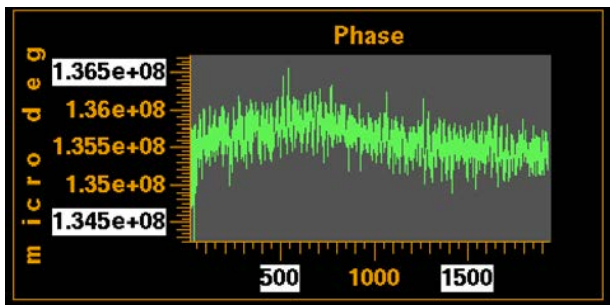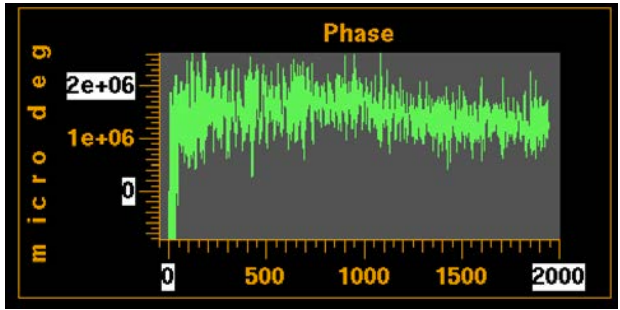

Figure 10: Beam RF phases measured at delta-t loops (in micro degrees).

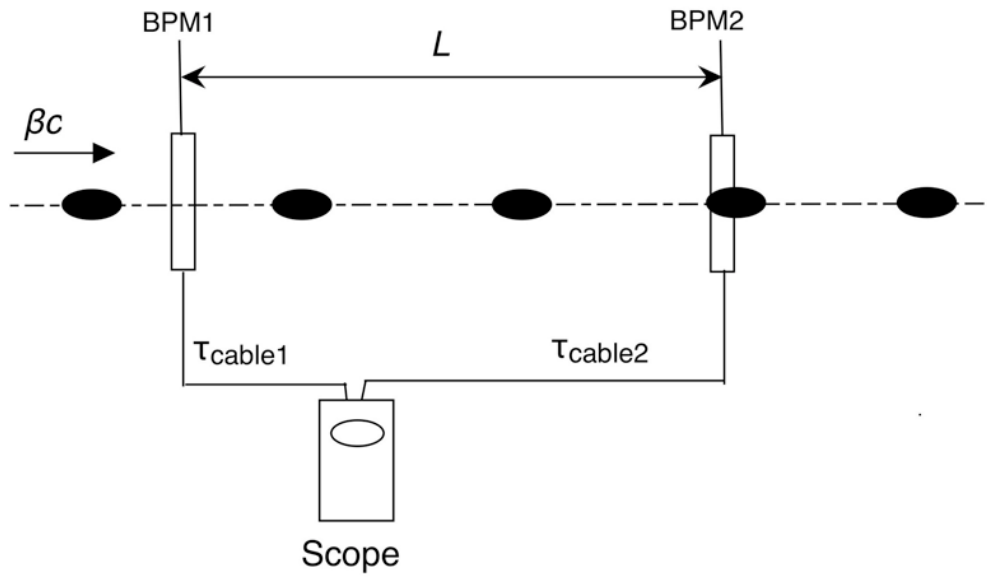

Figure 11: Time of flight measurement of absolute beam energy.

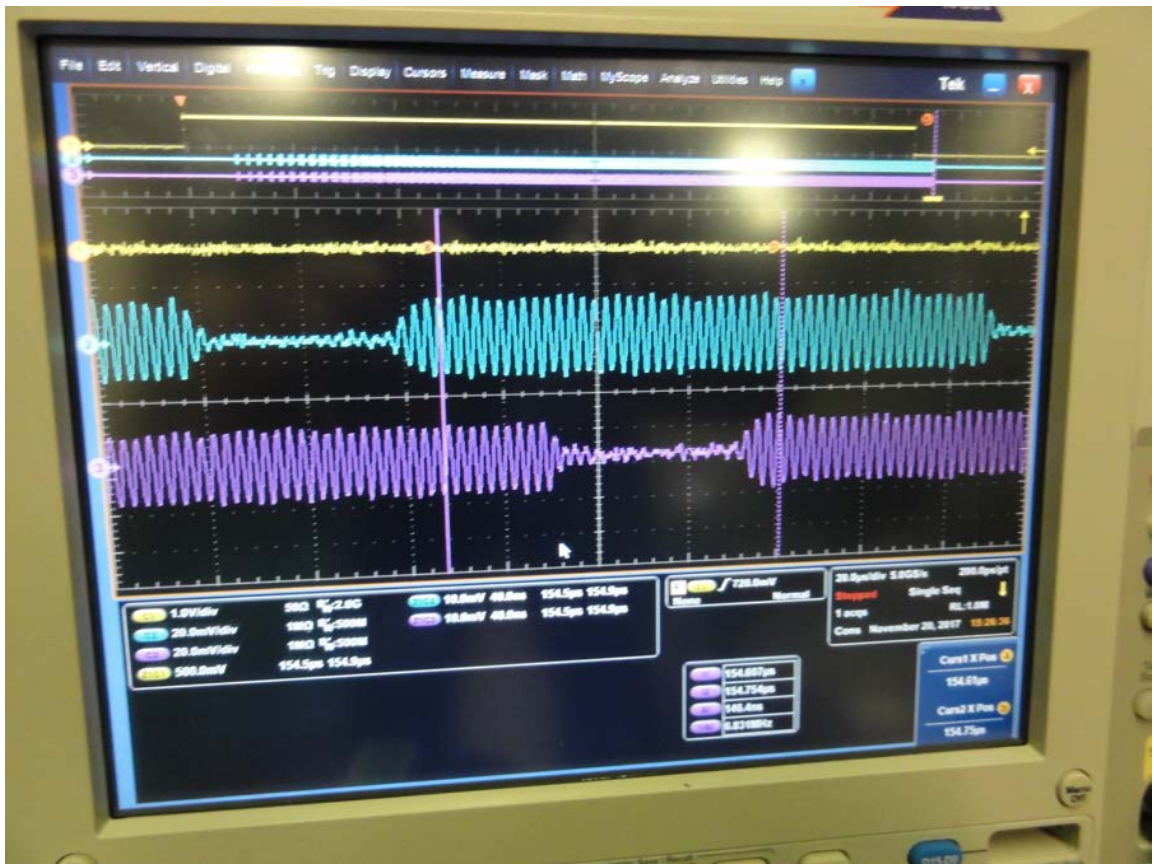

Figure 12: Measurement of time of flight of bunch train between two delta-t loops.

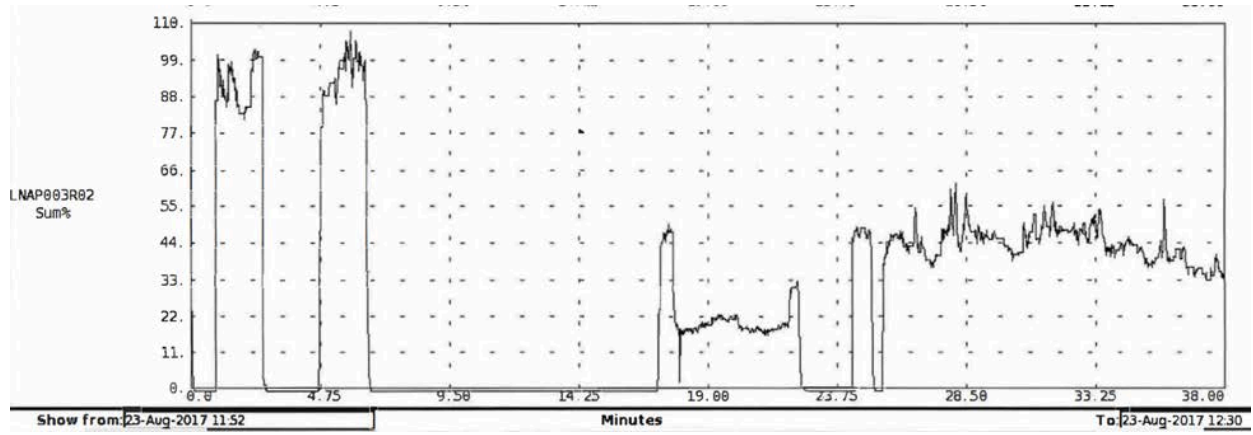

Figure 13: Reduction of sum beam losses in 805-MHz linac through change of quadrupole ramp.

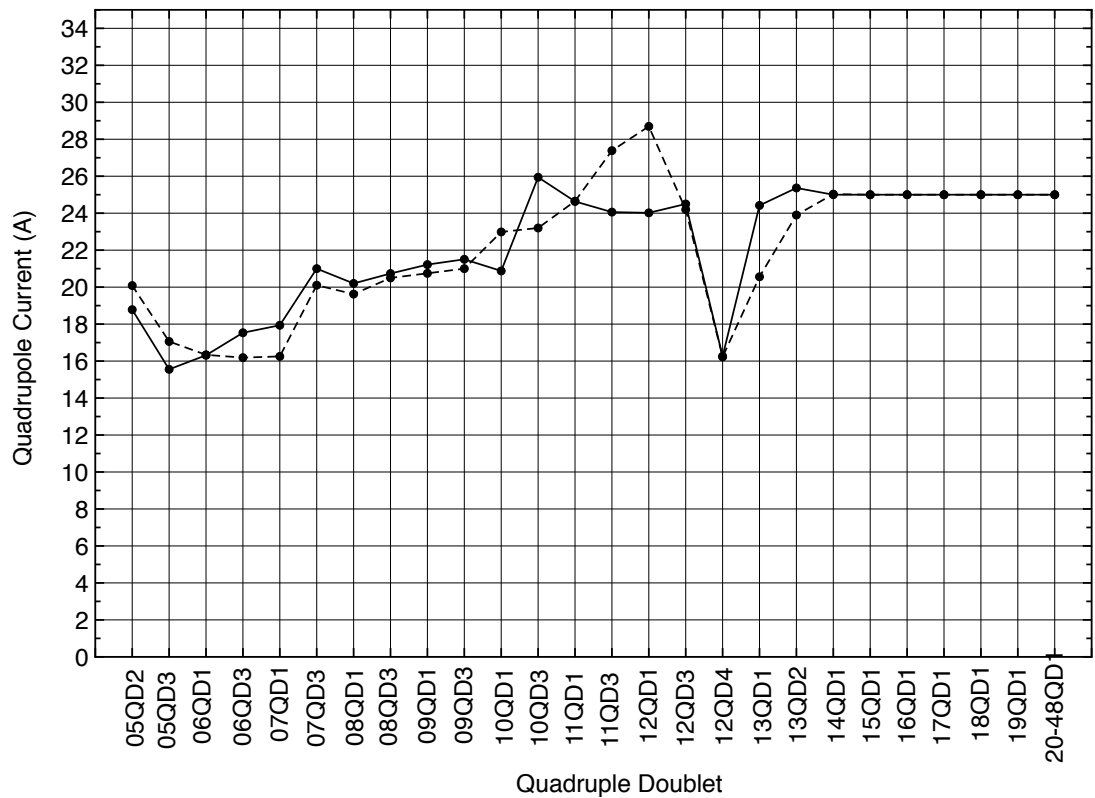

Figure 14: Quadruple setup in 805-MHz linac: (dotted) before, (solid) after reduction of beam losses.

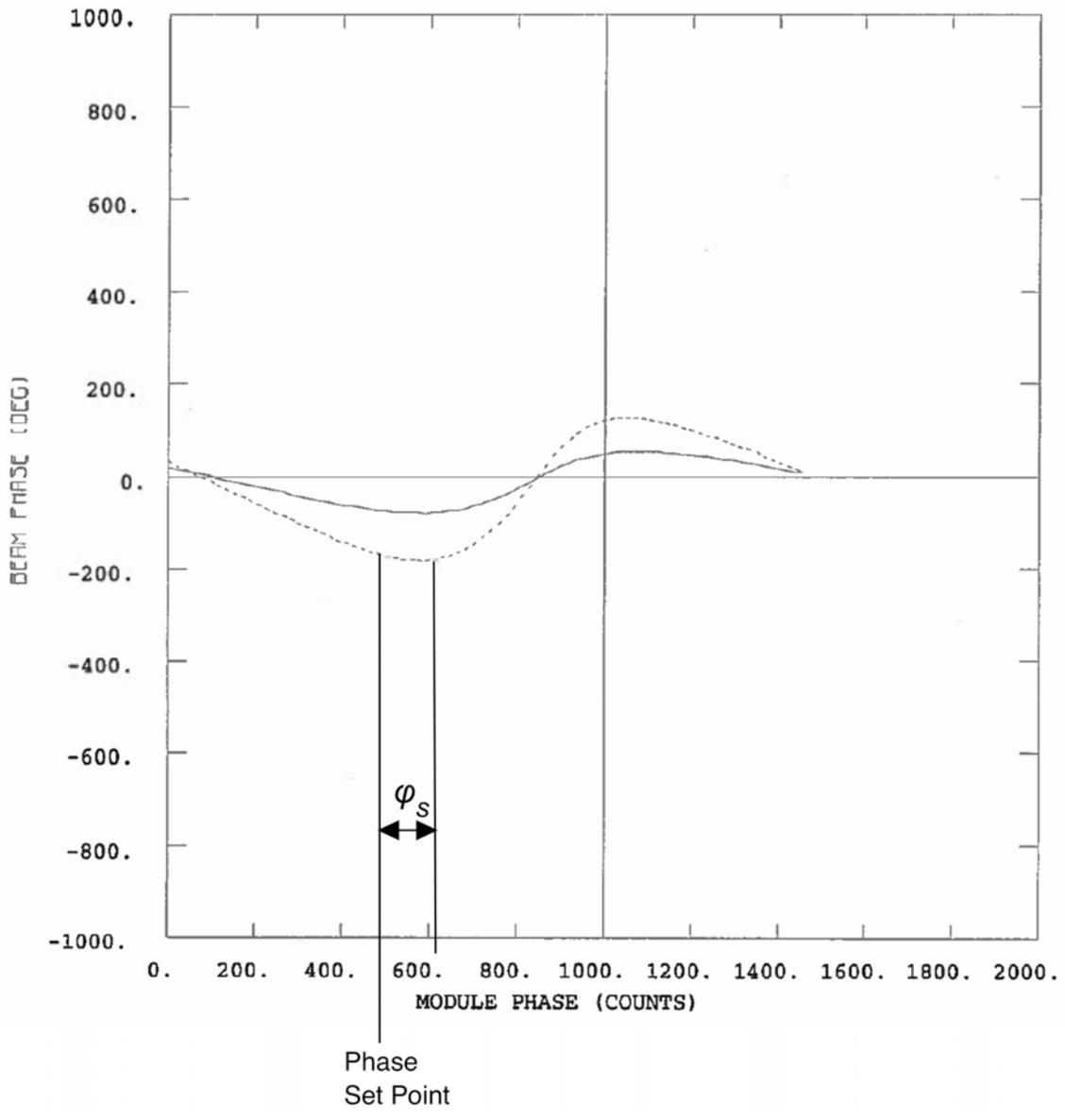

Figure 15: Phase scan.

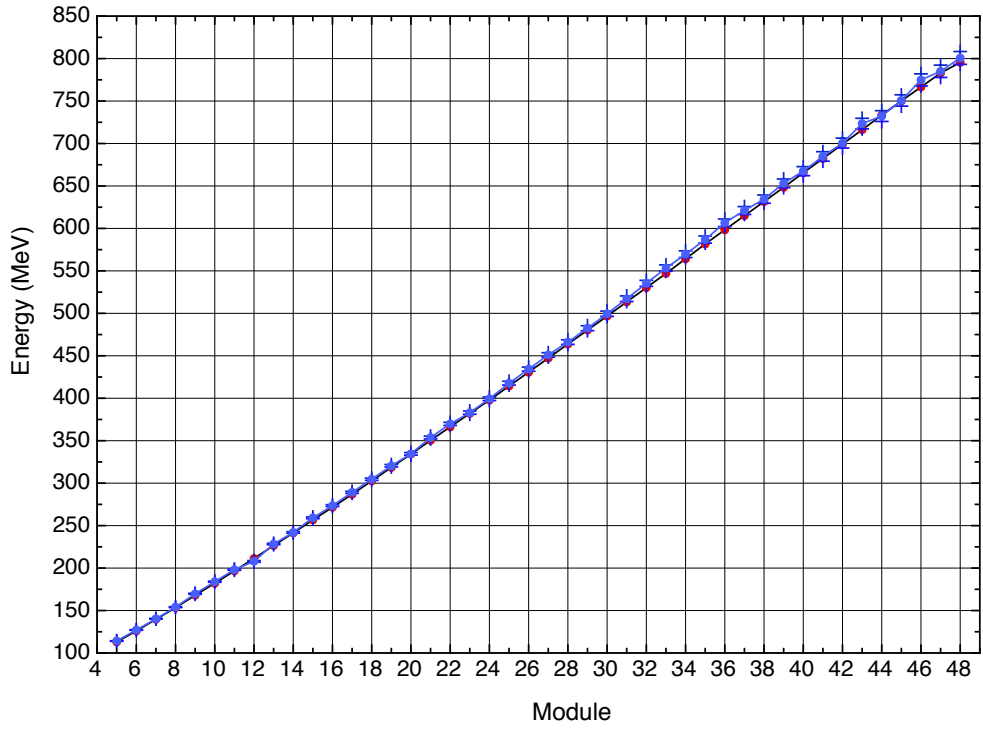

Figure 16: Results of beam energy measurement using absolute TOF method: (red) design, (blue) measured.

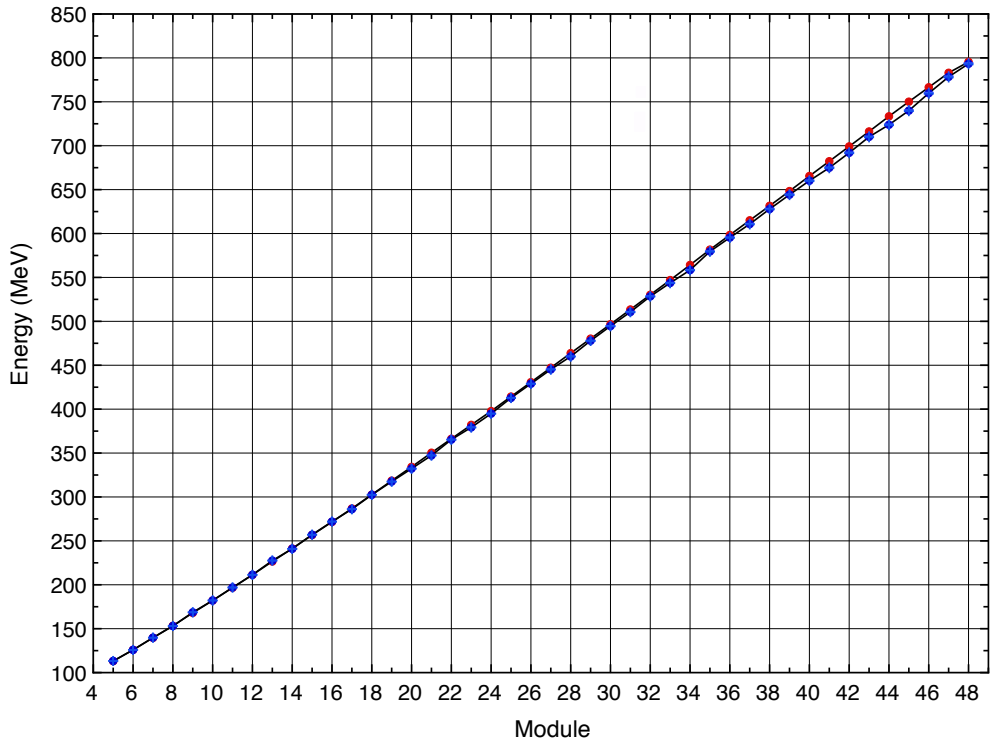

Figure 17: Results of beam energy measurement using RF phase difference method: : (red) design, (blue) measured.

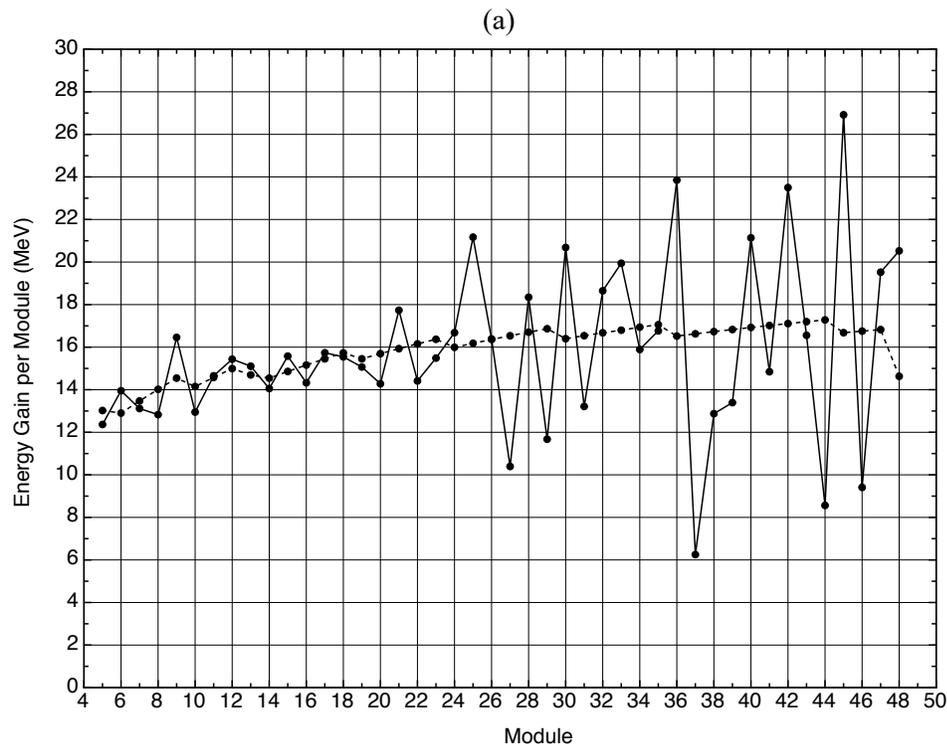

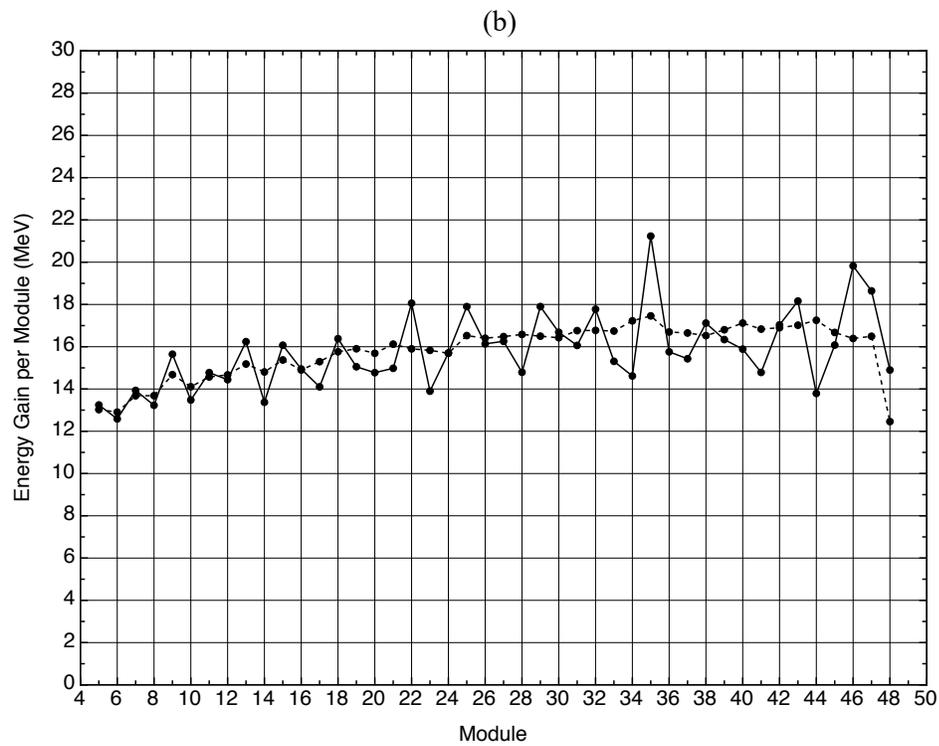

Figure 18: Energy gain per module: (dashed) design, (solid, a) after delta-t tuning, (solid, b) after improved tuning.

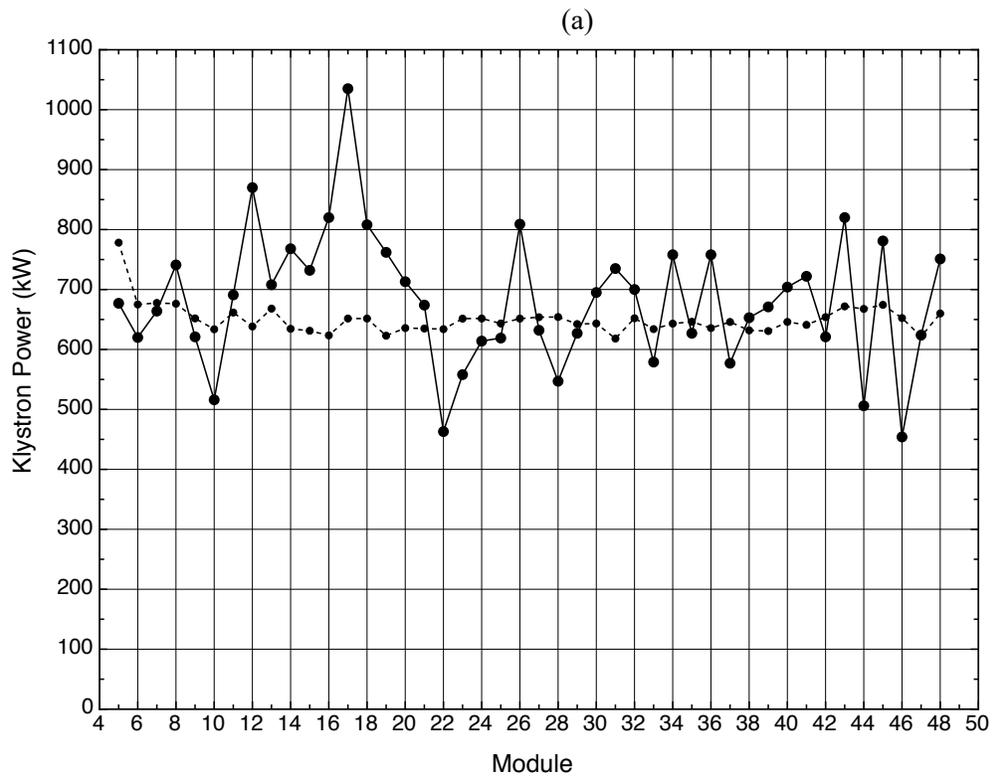
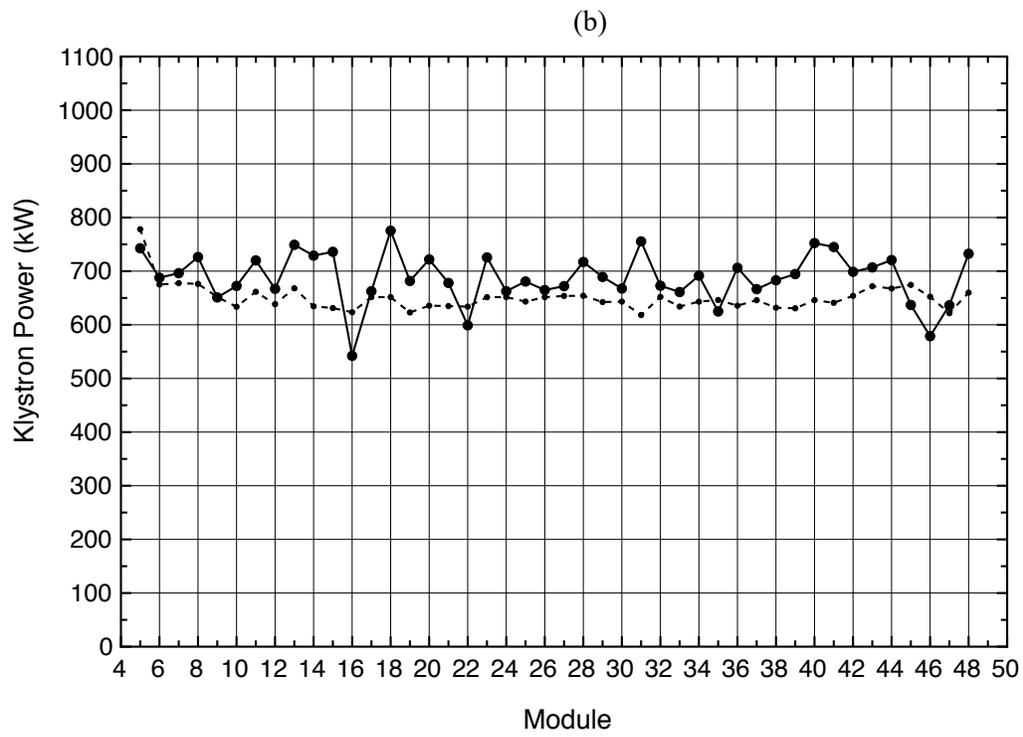

Figure 19: Klystron power: (dashed) design, (solid, a) after delta-t tuning, (solid, b) after improved tuning.

Table 1. Design parameters of 805-MHz linac.

| Module | Energy (MeV) | Phase (deg) | $R_{sh}$ (MOhm) | Length $L_a$ (m) | $UT$ (MV) | RF Power (kW) | Power Loss (kW) | Total Power (kW) |
|---|---|---|---|---|---|---|---|---|
| 5 | 113.02200 | -36.0 | 362.14945 | 11.79010 | 16.09360 | 715.18518 | 63.00000 | 778.18518 |
| 6 | 125.92400 | -35.5 | 401.82556 | 11.68780 | 15.84787 | 625.03479 | 50.00000 | 675.03479 |
| 7 | 139.39500 | -35.0 | 428.17035 | 12.20490 | 16.44505 | 631.61707 | 46.00000 | 677.61707 |
| 8 | 153.41299 | -34.5 | 461.23096 | 12.70280 | 17.00951 | 627.28552 | 49.00000 | 676.28552 |
| 9 | 167.95799 | -34.0 | 505.57632 | 13.18170 | 17.54443 | 608.82416 | 43.00000 | 651.82416 |
| 10 | 182.10600 | -33.5 | 493.20935 | 12.82690 | 16.96636 | 583.64117 | 50.00000 | 633.64117 |
| 11 | 196.68500 | -33.0 | 502.31564 | 13.21910 | 17.38345 | 601.58282 | 60.00000 | 661.58282 |
| 12 | 211.67900 | -32.5 | 539.08020 | 13.59650 | 17.77822 | 586.30444 | 52.00000 | 638.30444 |
| 13 | 226.37399 | -32.0 | 472.01819 | 13.29720 | 17.32801 | 636.11957 | 32.00000 | 668.11957 |
| 14 | 240.91701 | -31.5 | 480.51520 | 13.16380 | 17.05645 | 605.43842 | 29.00000 | 634.43842 |
| 15 | 255.77499 | -31.0 | 498.93219 | 13.45020 | 17.33382 | 602.20856 | 29.00000 | 631.20856 |
| 15 | 270.93399 | -30.5 | 521.58801 | 13.72600 | 17.59341 | 593.43390 | 30.00000 | 623.43390 |
| 17 | 286.38599 | -30.0 | 520.54889 | 13.99150 | 17.84242 | 611.57001 | 40.00000 | 651.57001 |
| 18 | 302.11700 | -30.0 | 544.88202 | 14.24700 | 18.16461 | 605.54950 | 46.00000 | 651.54950 |
| 19 | 317.56900 | -30.0 | 542.37988 | 13.99770 | 17.84242 | 586.95404 | 36.00000 | 622.95404 |
| 20 | 333.26401 | -30.0 | 543.15820 | 14.21880 | 18.12303 | 604.69330 | 31.00000 | 635.69330 |
| 21 | 349.19299 | -30.0 | 559.26099 | 14.43180 | 18.39320 | 604.92322 | 30.00000 | 634.92322 |
| 22 | 365.34699 | -30.0 | 575.23407 | 14.63700 | 18.65302 | 604.85828 | 29.00000 | 633.85828 |
| 23 | 381.71701 | -30.0 | 581.52026 | 14.83470 | 18.90247 | 614.42993 | 37.00000 | 651.42993 |
| 24 | 397.70902 | -30.0 | 560.24408 | 14.49470 | 18.46597 | 608.64905 | 43.00000 | 651.64905 |
| 25 | 413.88901 | -30.0 | 581.46790 | 14.66560 | 18.68304 | 600.30151 | 43.00000 | 643.30151 |
| 26 | 430.24899 | -30.0 | 579.82422 | 14.83040 | 18.89088 | 615.47150 | 36.00000 | 651.47150 |
| 27 | 446.78400 | -30.0 | 585.33411 | 14.98940 | 19.09297 | 622.79248 | 31.00000 | 653.79248 |
| 28 | 463.48700 | -30.0 | 598.87305 | 15.14270 | 19.28696 | 621.14490 | 33.00000 | 654.14490 |
| 29 | 480.35199 | -30.0 | 619.29150 | 15.29070 | 19.47401 | 612.37231 | 30.00000 | 642.37231 |
| 30 | 496.75000 | -30.0 | 593.30102 | 14.86970 | 18.93479 | 604.29040 | 39.00000 | 643.29040 |
| 31 | 513.28802 | -30.0 | 632.90295 | 14.99770 | 19.09646 | 576.19391 | 42.00000 | 618.19391 |
| 32 | 529.96301 | -30.0 | 601.83167 | 15.12140 | 19.25461 | 616.01947 | 36.00000 | 652.01947 |
| 33 | 546.76001 | -30.0 | 621.82458 | 15.24080 | 19.39550 | 604.97022 | 29.00000 | 633.97022 |
| 34 | 563.69898 | -30.0 | 625.00549 | 15.35640 | 19.55943 | 612.10846 | 31.00000 | 643.10846 |
| 35 | 580.75299 | -30.0 | 630.31964 | 15.46800 | 19.69228 | 615.22076 | 31.00000 | 646.22076 |
| 36 | 597.27399 | -30.0 | 609.93835 | 14.98620 | 19.07680 | 596.65735 | 39.00000 | 635.65735 |
| 37 | 613.90002 | -30.0 | 612.36169 | 15.08280 | 19.19809 | 601.87738 | 44.00000 | 645.87738 |
| 38 | 630.63000 | -30.0 | 625.28565 | 15.17650 | 19.31811 | 596.83038 | 35.00000 | 631.83038 |
| 39 | 647.45801 | -30.0 | 627.49121 | 15.26710 | 19.43130 | 601.72211 | 29.00000 | 630.72211 |
| 40 | 664.38300 | -30.0 | 621.10864 | 15.35500 | 19.54329 | 614.93280 | 31.00000 | 645.93280 |
| 41 | 681.40100 | -30.0 | 633.02741 | 15.44000 | 19.65070 | 610.00494 | 31.00000 | 641.00494 |
| 42 | 698.50897 | -30.0 | 635.62915 | 15.52240 | 19.75458 | 613.94836 | 40.00000 | 653.94836 |
| 43 | 715.70502 | -30.0 | 627.96796 | 15.60220 | 19.85628 | 627.85321 | 44.00000 | 671.85321 |
| 44 | 732.98499 | -30.0 | 630.31995 | 15.67960 | 19.95318 | 631.63086 | 36.00000 | 667.63086 |
| 45 | 749.66400 | -30.0 | 577.37665 | 15.13530 | 19.25926 | 642.42163 | 32.00000 | 674.42163 |
| 46 | 766.41699 | -30.0 | 601.26794 | 15.20250 | 19.34468 | 622.37927 | 30.00000 | 652.37927 |
| 47 | 783.24103 | -30.0 | 638.96210 | 15.26770 | 19.42672 | 590.64124 | 31.00000 | 621.64124 |
| 48 | 795.46130 | -30.0 | 613.23999 | 15.33100 | 14.11076 | 324.69092 | 41.00000 | 660.0 |

Table 2. Results of beam energy measurement along linac (MeV).

| Module | Design | Absolute Method | Error | RF Phase Method | Error |
|---|---|---|---|---|---|
| 5 | 113.02200 | 114.08083 | 0.30722 | 113.24384 | 0.04476 |
| 6 | 125.92400 | 127.08555 | 0.36473 | 125.83000 | 0.05264 |
| 7 | 139.39500 | 140.33302 | 0.42737 | 139.76276 | 0.06197 |
| 8 | 153.41299 | 153.99704 | 0.49624 | 152.99019 | 0.07140 |
| 9 | 167.95799 | 169.73883 | 0.58089 | 168.63522 | 0.08327 |
| 10 | 182.10600 | 183.98119 | 0.66232 | 182.11415 | 0.09414 |
| 11 | 196.68500 | 198.15321 | 0.74791 | 196.88849 | 0.10667 |
| 12 | 211.67900 | 207.87032 | 0.92090 | 211.32892 | 0.13326 |
| 13 | 226.37399 | 228.35078 | 0.80131 | 227.56898 | 0.10868 |
| 14 | 240.91701 | 242.01938 | 0.88281 | 240.94121 | 0.12795 |
| 15 | 255.77499 | 259.19778 | 0.99033 | 257.01297 | 0.13148 |
| 16 | 270.93399 | 273.55109 | 1.08452 | 271.95874 | 0.15346 |
| 17 | 286.00000 | 289.21753 | 1.19180 | 286.06113 | 0.16699 |
| 18 | 302.11700 | 304.62326 | 1.30209 | 302.44037 | 0.18338 |
| 19 | 317.56900 | 320.48016 | 1.42038 | 317.49307 | 0.19909 |
| 20 | 333.26401 | 334.54123 | 1.52940 | 332.26572 | 0.21511 |
| 21 | 349.19299 | 353.70993 | 1.68434 | 347.24509 | 0.23197 |
| 22 | 365.34698 | 369.79501 | 1.81999 | 365.30392 | 0.25312 |
| 23 | 381.71701 | 383.05975 | 1.93578 | 379.20084 | 0.27003 |
| 24 | 397.70901 | 399.38010 | 2.08311 | 394.90634 | 0.28978 |
| 25 | 413.88901 | 417.76297 | 2.25558 | 412.80426 | 0.31315 |
| 26 | 430.24899 | 434.26181 | 2.41629 | 428.94019 | 0.33501 |
| 27 | 446.78400 | 451.04346 | 2.58556 | 445.19659 | 0.35779 |
| 28 | 463.48700 | 466.09918 | 2.74240 | 459.98828 | 0.37919 |
| 29 | 480.35199 | 482.49545 | 2.91874 | 477.88779 | 0.40595 |
| 30 | 496.75000 | 499.52435 | 3.10790 | 494.57202 | 0.43175 |
| 31 | 513.28802 | 517.20819 | 3.31000 | 510.63348 | 0.45738 |
| 32 | 529.96301 | 535.31976 | 3.52596 | 528.40881 | 0.48665 |
| 33 | 546.76001 | 553.30121 | 3.74655 | 543.71771 | 0.51263 |
| 34 | 563.69897 | 569.62103 | 3.95296 | 558.33362 | 0.53812 |
| 35 | 580.75299 | 586.85370 | 4.17739 | 579.56799 | 0.57632 |
| 36 | 597.27399 | 606.64679 | 4.44340 | 595.32568 | 0.60559 |
| 37 | 613.90002 | 621.09338 | 4.64300 | 610.75647 | 0.63502 |
| 38 | 630.63000 | 634.48639 | 4.83284 | 627.87476 | 0.66856 |
| 39 | 647.45801 | 653.08954 | 5.10305 | 644.21381 | 0.70146 |
| 40 | 664.38300 | 667.44843 | 5.31717 | 660.09955 | 0.73428 |
| 41 | 681.40100 | 684.77600 | 5.58200 | 674.88367 | 0.76556 |
| 42 | 698.50897 | 700.49951 | 5.82860 | 691.91602 | 0.80252 |
| 43 | 715.70502 | 723.44116 | 6.24200 | 710.07696 | 0.84298 |
| 44 | 732.98499 | 732.30804 | 6.34560 | 723.86743 | 0.87446 |
| 45 | 749.66400 | 750.68152 | 6.65500 | 739.94501 | 0.91049 |
| 46 | 766.41699 | 774.93665 | 7.07751 | 759.76819 | 0.95946 |
| 47 | 783.24103 | 785.06848 | 7.25800 | 778.41003 | 1.00484 |
| 48 | 795.46002 | 800.75824 | 7.54292 | 793.30792 | 1.04292 |

Table 3. Beam spill in high-energy beamlines before and after optimization of tuning.

| Activation Protection Device | Beam spill before tuning optimization (nA) | Beam spill after tuning optimization (nA) |
|---|---|---|
| LAAP1 | 1.0 | 0.7 |
| LAAP2 | 2.0 | 1.8 |
| LAAP3 | 1.0 | 0.15 |
| LAAP4 | 2.0 | 12.8 |
| LAAP5 | 2.0 | 1.35 |
| XDAP1 | 60.0 | 0.95 |
| XDAP2 | 40.0 | 0.1 |
| XDAP3 | 5.0 | 0.0 |
| XDAP4 | 15.0 | 5.4 |
| XDAP5 | 50.0 | 15.7 |
| XDAP6 | 10.0 | 2.15 |
| XDAP7 | 80.0 | 39.0 |
| LDAP1 | 5.0 | 2.0 |
| LDAP2 | 40.0 | 80.0 |
| LDAP3 | 90.0 | 5.0 |
| LDAP4 | 5.0 | 5.0 |
| LDAP5 | 25.8 | 40.0 |
| LDAP6 | 5.0 | 15.0 |
| LDAP7 | 4.0 | 7.0 |
| LDAP8 | 3.0 | 7.0 |
| LDAP9 | 5.0 | 10.0 |
| LDAP10 | 0.0 | 0.0 |
|  |  |  |
| Total | 450.8 | 253.0 |